\documentclass[sigconf]{acmart} 
\AtBeginDocument{%
  \providecommand\BibTeX{{%
    \normalfont B\kern-0.5em{\scshape i\kern-0.25em b}\kern-0.8em\TeX}}}

\copyrightyear{2023}
\acmYear{2023}
\setcopyright{rightsretained}
\acmConference[CHI EA '23]{Extended Abstracts of the 2023 CHI Conference on Human Factors in Computing Systems}{April 23--28, 2023}{Hamburg, Germany}
\acmBooktitle{Extended Abstracts of the 2023 CHI Conference on Human Factors in Computing Systems (CHI EA '23), April 23--28, 2023, Hamburg, Germany}\acmDOI{10.1145/3544549.3577064}
\acmISBN{978-1-4503-9422-2/23/04}

\begin{document}

\title{Understanding Physical Breakdowns in Virtual Reality}

\author{Wen-Jie Tseng}
\affiliation{%
  \institution{LTCI, INFRES, Telecom Paris, IP Paris}
  \city{Palaiseau}
  \state{}
  \country{France}
}
\email{wen-jie.tseng@telecom-paris.fr}

\renewcommand{\shortauthors}{Wen-Jie Tseng}
\newcommand{\VRSB}{SB}

\begin{abstract}
Virtual Reality (VR) moves away from well-controlled laboratory environments into public and personal spaces. As users are visually disconnected from the physical environment, interacting in an uncontrolled space frequently leads to collisions and raises safety concerns. In my thesis, I investigate this phenomenon which I define as \emph{the physical breakdown in VR}. The goal is to understand the reasons for physical breakdowns, provide solutions, and explore future mechanisms that could perpetuate safety risks. First, I explored the reasons for physical breakdowns by investigating how people interact with the current VR safety mechanism (e.g., Oculus Guardian). Results show one reason for breaking out of the safety boundary is when interacting with large motions (e.g., swinging arms), the user does not have enough time to react although they see the safety boundary. I proposed a solution, FingerMapper, that maps small-scale finger motions onto virtual arms and hands to enable whole-body virtual arm motions in VR to avoid physical breakdowns. To demonstrate future safety risks, I explored the malicious use of perceptual manipulations (e.g., redirection techniques) in VR, which could deliberately create physical breakdowns without users noticing. Results indicate further open challenges about the cognitive process of how users comprehend their physical environment when they are blindfolded in VR. 
\end{abstract}



\maketitle

\section{Introduction}
Virtual Reality (VR) becomes more available in public and personal spaces, eliciting new contexts, like social interaction \cite{gugenheimerShareVREnablingCoLocated2017}, in-vehicle \cite{mcgillAmPassengerHow2017}, and confined space \cite{tsengFingerMapperEnablingArm2021, liRearSeatProductivityVirtual2021}. These contexts all share one common dilemma --- users are visually disconnected from the physical environment. Perhaps this outcome might originate from pursuing a higher presence and better immersive experience. However, imagine using VR in an uncertain environment (e.g., home). Accidents could happen and did happen. Prominent examples would be colliding, hitting, and falling over, identified in people's everyday VR usage \cite{daoBadBreakdownsUseful2021}. I define these accidents as \emph{the physical breakdown in VR} --- an abrupt disruption of the VR experience caused by a collision with the physical environment.

Physical breakdowns become relevant because VR is available in uncontrolled environments. They deviate the VR user's attention from the virtual environment to the real world and lead to safety concerns that people may get hurt when it occurs. Although current VR Head-Mounted Displays (HMDs) provide safety mechanisms (e.g., Oculus Guardian) to avoid these accidents, they still cannot cover all types of physical breakdowns in VR. There is a lack of knowledge about the process of physical breakdowns in VR and how future technologies should mitigate them.

\begin{figure*}[t]
\begin{center}
  \begin{tabular}{@{\hspace{0.1cm}}c}
		\includegraphics[width=\linewidth]{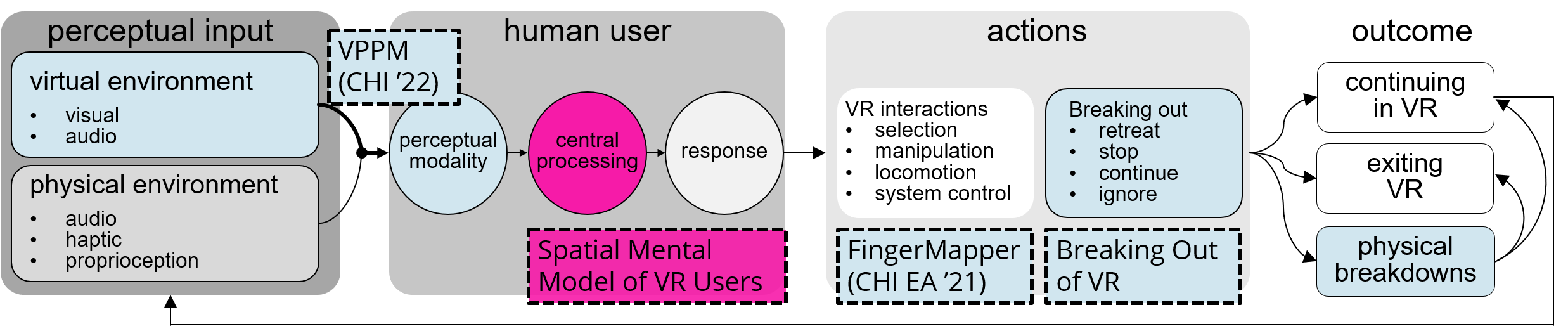}
  \end{tabular}
  \caption{My research explores physical breakdowns in VR within a structure adapted from the attentional model for synthetic environment \cite{draperTelepresence1998}, in which I position my research projects (sky-blue) and future directions (pink).}
\label{fig:thesis-structure}
\Description{Figure 1 shows my thesis structure adapted from the attentional model for synthetic environments. A human user perceives the perceptual input from the virtual and physical environment and then performs actions. I position physical breakdowns in VR as an outcome of the VR user's action within the structure. My current research progress focuses on actions. For example, I investigate why VR users break out of the safety boundary and develop solutions (FingerMapper) to reduce the required interaction space to avoid physical breakdown. I also explore how future physical breakdowns could be exploited by malicious actors using perceptual manipulations. Future research aims to understand the behavioral process of the human user when a physical breakdown happens, to inform the future safety mechanisms in VR.}
\end{center}
\end{figure*}

My thesis investigates the reasons for physical breakdowns, how they could happen, and how to mitigate them by exploring solutions and future safety mechanisms. I position physical breakdowns into the attentional model for synthetic environment \cite{draperTelepresence1998} by adding physical breakdowns as an outcome of actions (Figure \ref{fig:thesis-structure}). The model interprets how people elicit a sense of presence in VR. A VR user perceives and processes all the sensory information from the virtual and physical environments. Presence is a psychological state arising from the commitment of attentional resources to the virtual environment. The VR user then makes an action to interact in VR (e.g., interaction or locomotion). These actions lead to several outcomes, like continuing or exiting VR. I explore physical breakdowns in VR within this structure and present the following research questions.

\begin{itemize}
    \item RQ1: How and why physical breakdowns happen with the current HMD-based VR technology? 
    \item RQ2: What harm and risks come with the physical breakdown, and how to mitigate them?
    \item RQ3: What are the behavior and cognitive process behind the physical breakdown phenomena?
\end{itemize}

To address RQ1, I started by investigating empirical evidence of physical breakdown by observing how users perceive and interact with the safety boundary in their everyday VR usage through an online survey and a lab study. Results indicate VR users break out of the safety boundary for accidental and intentional reasons. One accidental reason was that participants could not stop themselves while performing large gestures, although they saw the safety boundary. Therefore, I developed a solution, FingerMapper, which maps small-scale finger motions onto virtual arms and hands to enable whole-body virtual movements when using VR in confined spaces. This technique mitigates collisions with the physical environment by reducing the interaction space, contributing to RQ2. FingerMapper shows that the virtual and physical movements do not need to keep a one-to-one mapping. Perceptual manipulations in VR (e.g., redirection techniques) leverage this discrepancy to direct the user's actions. I explored how malicious actors could exploit perceptual manipulations to provoke physical breakdowns through a speculative design workshop. By analyzing scenarios created by our participants, I identified two main risks (puppetry and mismatching) and proposed suggestions and mitigations for current practitioners and the research committee (RQ2). Finally, all research projects point to open challenges in RQ3. For example, when identifying actions and reasons when interacting with safety boundaries, I found participants sometimes rely on their understanding of the physical environment when deciding whether to break out. My thesis aims to understand the psychological process of physical breakdowns with empirical studies and inform future mechanisms that mitigate risks and safety concerns.

\paragraph{Research Situation} 
I am in the third year of my PhD at Telecom Paris, France. In 2023, I will transition to the Department of Computer Science, TU Darmstadt, Germany, to continue my PhD. I plan to hand in my thesis in approximately 2.5 years. By participating in the Doctoral Consortium, I hope to get valuable feedback on my research direction and approach.

\section{Background and Key Related Work}

\paragraph{Virtual Reality and Presence}
VR technologies immerse users by providing multi-sensory input (e.g., visual, audio, haptic) and motion tracking, enabling 3D interaction inside a simulated environment. A user feels a psychological state of ``being there'' inside a computer-mediated environment \cite{witmerMeasuringPresenceVirtual1998, slaterDepthPresenceVirtual1994} called the sense of presence. However, a VR user may be distracted by stimuli from the physical environment (e.g., break in presence \cite{slaterVirtualPresenceCounter2000}). This phenomenon can be explained using the attentional model \cite{draperTelepresence1998} where participants' attention resources shift from the virtual content to the physical environment. These findings indicate we are not always present inside the virtual environment. Instead, multiple transitions happen within the virtual/physical environments and perceptual/conceptual tasks. The physical breakdown is an abrupt transition in the context of presence. Studying the physical breakdown in VR improves the user's safety and opens opportunities to explore the behavior process behind these transitions.


\paragraph{Physical Breakdowns as Potential Harm and Risks in VR}
A breakdown represents the moment of an abrupt disruption of an experience \cite{bodkerApplyingActivityTheory1995}. Dao and colleagues \cite{daoBadBreakdownsUseful2021} applied the breakdown concept to explore VR fails (e.g., overreacting, colliding, hitting) and their causes by analyzing YouTube videos. Here, I extend their work and define ``the physical breakdown in VR'' as \textit{an abrupt disruption of the VR experience caused by a collision with the physical environment}. Physical breakdowns are a subset of VR fails, which could potentially induce harm to the user. In some cases, they may lead to exiting a VR experience \cite{knibbeDreamCollapsingExperience2018}.

The safety boundary (e.g., Oculus Guardian, Vive Chaperone) can prevent accidents in VR by displaying 2D grids based on the user's proximity. Recent research explored multiple sensory feedback to notify the VR user \cite{georgeInvisibleBoundariesVR2020, faltaousSaVRIncreasingSafety2020}. Nevertheless, the physical environment is usually unknown and uncontrolled at home. Blending the virtual content with the physical surroundings can keep the presence in VR and prevent collision \cite{hartmannRealityCheckBlendingVirtual2019}. Still, blending techniques are not as prevalent as safety boundaries. Instead of building new artifacts, I study the VR user's behavior while interacting with safety boundaries to understand the behavior process before physical breakdowns happen and inform future safety mechanisms.

VR technologies are highly persuasive for benefits (e.g., training) as well as malicious purposes because of the realism \cite{slaterEthicsRealismVirtual2020}. Interaction techniques in VR leverage the visual limits of the user (e.g., humans are visually dominant when combining several sources of sensory information \cite{posnerVisualDominanceInformationprocessing1976, vanbeersWhenFeelingMore2002}) to hack human perception. Examples like redirection techniques \cite{razzaqueRedirectedWalkingPlace2002, azmandianHapticRetargetingDynamic2016, sunVirtualRealityInfinite2018} and pseudo haptic illusions \cite{lecuyerPseudohapticFeedbackCan2000, rietzlerBreakingTrackingEnabling2018}, can direct the user's action to overcome limitations in current VR systems (e.g., limited tracked space, lack of haptic feedback). I define them as \emph{Virtual-Physical Perceptual Manipulation} (VPPM) referring to Extended Reality (XR) driven exploits that \emph{alter the human multi-sensory perception of our physical actions and reactions to nudge the user's physical movements} (e.g.,  the position of body and hands). While VPPMs had positive intents, malicious actors can also exploit VPPMs to provoke physical breakdowns and even harm the VR user. Casey et al. identified the \emph{human joystick attack} directing an immersed user's physical movement to a location without the user's knowledge \cite{caseyImmersiveVirtualReality2019}. My thesis broadly explores how malicious actors could impose VPPMs to manipulate the user's body movement and provoke future physical breakdowns using a speculative design workshop.


\section{Dissertation Status}

\subsection{Breaking Out of the Safety Boundary in VR}
In this project, I investigated physical breakdowns in the everyday usage of VR to address RQ1. Instead of observing physical breakdowns directly, I chose the moment when VR users break out of the Safety Boundary (\VRSB). Break-outs happen during interaction with {\VRSB}s, and they are one step before a physical breakdown could happen (e.g., colliding with objects in the environment). I explored how VR users interact with {\VRSB}s and the reasons for break-outs in VR (RQ1 and RQ3) using an online survey (n=64) and a lab study (n=12) with a mixed-method approach. 

\begin{figure}[t]
\begin{center}
  \begin{tabular}{@{\hspace{0.1cm}}c}
		\includegraphics[width=\linewidth]{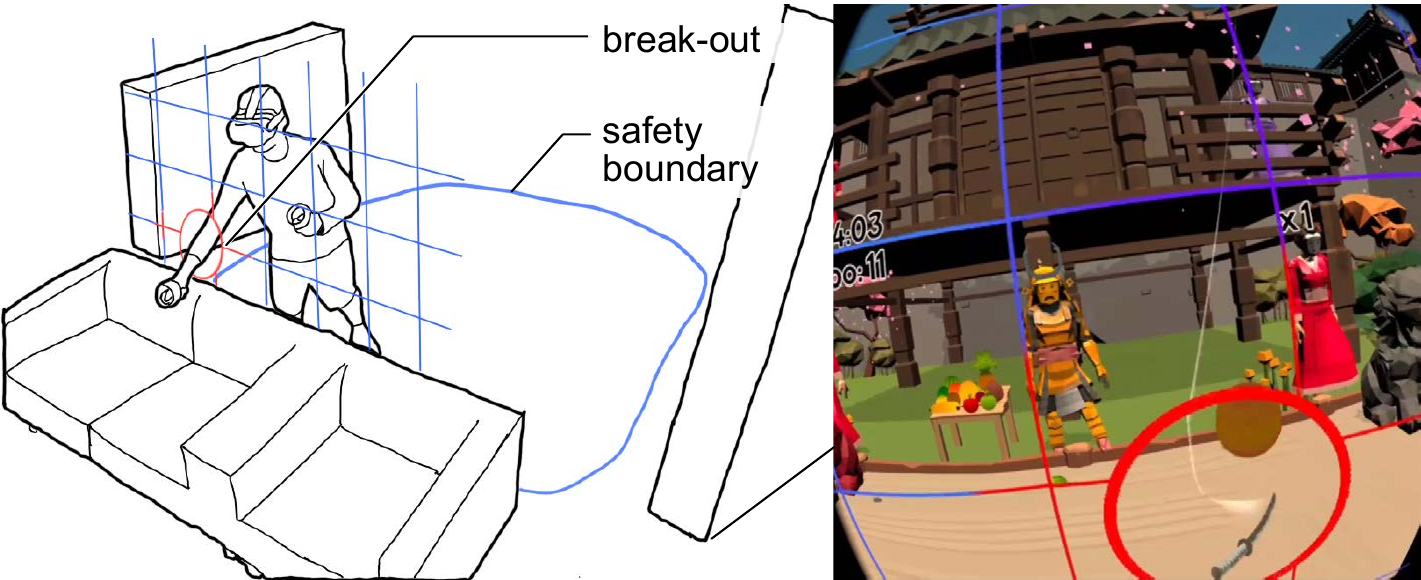}
  \end{tabular}
  \caption{Despite having the safety boundary, VR users may still break out accidentally or intentionally (the red circle), leading to safety concerns (e.g., collision). I implemented an apparatus (FruitSlicer) to replicate the break-out experience.}
\label{fig:bo}
\Description{Figure 2 shows a VR user who broke out of the safety boundary while playing a VR game. Although there is a safety boundary, VR users still break out of it for accidental or intentional reasons.}
\end{center}
\end{figure}

The online survey collected data about attitudes towards the {\VRSB}, behavior while interacting with {\VRSB}s, and reasons for breaking out. The results show participants perceived {\VRSB}s as positive and helpful, keeping them safe and alert during a VR experience. However, half of them (52\%) still broke out while using VR, indicating {\VRSB}s still cannot prevent participants from break-outs every time. By analyzing open-ended questions using a thematic analysis \cite{braunUsingThematicAnalysis2006}, I identified four physical behaviors (\textit{retreat}, \textit{stop}, \textit{adapt}, and \textit{ignore} in Figure \ref{fig:thesis-structure}) that VR users did when seeing {\VRSB}s. Further, I categorized multiple reasons for breaking out into two main classes of break-outs: intentional (57\%) and accidental (43\%). To understand intentional break-outs, I implemented FruitSlicer (Figure \ref{fig:bo}), a VR application that can provoke breaking out of \VRSB, allowing us to observe participants' reactions and interview them afterward about their reasons. The lab study exposed participants to break-out experiences with multiple obstacles in a mock-up living room (Figure \ref{fig:bo}). The analysis of the semi-structured interview revealed three interaction strategies. \textit{Cage} indicates participants tend to break out with arms instead of the body. \textit{Confine} means participants stay inside the \VRSB\ and avoid break-outs. Lastly, participants sometimes touched the obstacles, obtaining more spatial information about the physical environment (\textit{update}), and dared to break out intentionally because they assumed it would be safe. 

Through the two studies, I observed that intentional break-outs are usually associated with the participant's understanding of the virtual and physical environment. I interpret this observation with the Spatial Mental Model (SMM) \cite{leeInterplayVisualSpatial2005,franklinSwitchingPointsView1992} that VR user forms the spatial relation inside a small or well-known environment, and they recall their SMM when they decide whether to break out of the \VRSB\ (RQ3). This finding point to new open challenges in the user's behavior and cognitive process while interacting with VR and being aware of the physical environment simultaneously.

\begin{figure}[t]
\begin{center}
  \begin{tabular}{@{\hspace{0.1cm}}c}
		\includegraphics[width=\linewidth]{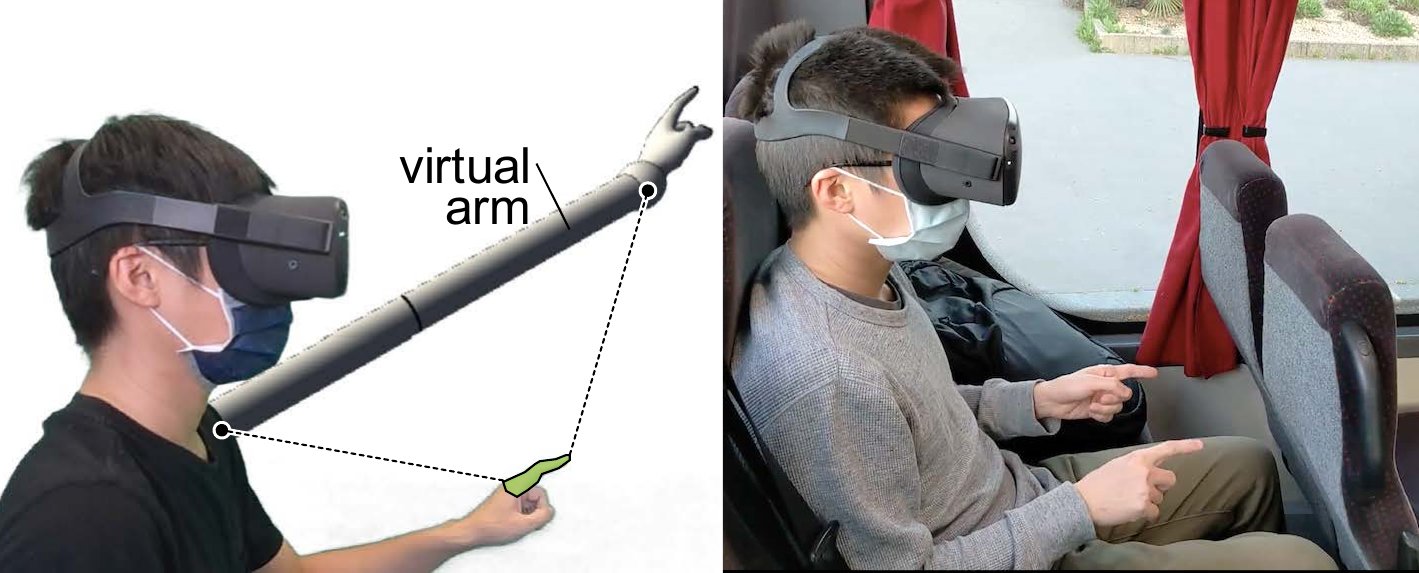}
  \end{tabular}
  \caption{FingerMapper leverages less physically demanding finger motions, enabling whole-body virtual arm movements in confined spaces (e.g., the passenger seat of a car) with fewer collisions while preserving presence and enjoyment.}
\label{fig:fm}
\Description{Figure 3 illustrates FingerMapper, a technique that maps small and less physically demanding motions of fingers onto the virtual arm. FingerMapper enables whole-body virtual arm interaction in confined spaces to avoid physical breakdowns in the mobile VR context, providing a higher perceived safety.}
\end{center}
\end{figure}

\subsection{FingerMapper [CHI EA '21]}
I found some participants broke out of safety boundaries because they had large movements. Although whole-body movements enhance the presence and enjoyment of VR experiences, using large gestures is often uncomfortable and impossible in confined spaces (e.g., public transport). This context may lead to collisions and break the VR experience.

I developed a solution to address RQ2, FingerMapper (Figure \ref{fig:fm}), through a user-centered design approach. FingerMapper maps small-scale finger motions onto virtual arms and hands to enable whole-body virtual movements in VR. Since the user has fewer physical movements, this technique can also reduce collisions. In a first target selection study (n=13) comparing FingerMapper to hand tracking and ray-casting, I found that FingerMapper can significantly reduce physical motions and fatigue while having a similar degree of precision. In a consecutive study (n=13), I compared FingerMapper to hand tracking inside a confined space (car). The results showed participants had significantly higher perceived safety and fewer collisions with FingerMapper while preserving a similar degree of presence and enjoyment as hand tracking. Finally, I present three applications demonstrating how FingerMapper could be applied for locomotion and interaction for VR in confined spaces. The results point out that in a different context (e.g., confined spaces), perceived safety should also be considered as other metrics like presence.

\subsection{Exploring Future Physical Breakdowns by Exploiting Perceptual Manipulations in VR [CHI '22]}
\begin{figure*}[t]
\begin{center}
  \begin{tabular}{@{\hspace{0.1cm}}c}
		\includegraphics[width=\linewidth]{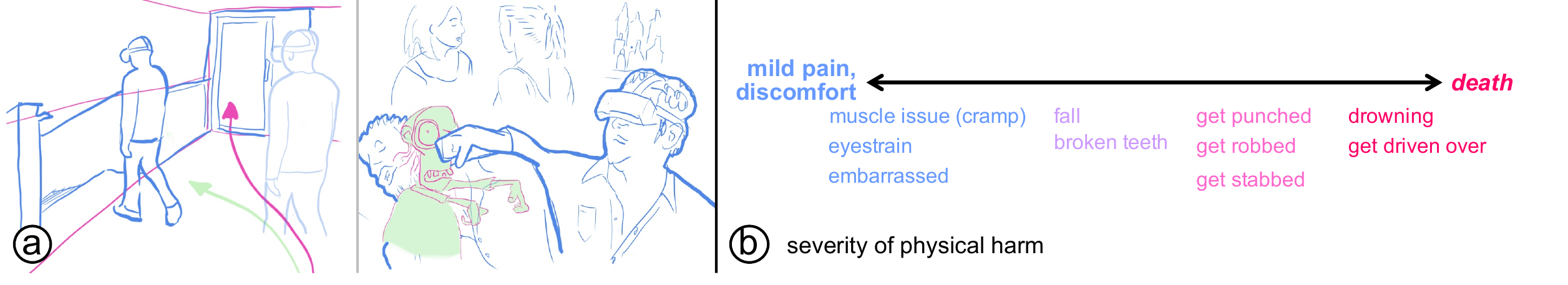}
  \end{tabular}
  \caption{(a) Participants created scenarios in the speculative design workshop to speculate on the malicious use of VPPMs. The left shows the puppetry attack that directs a user to fall over the stairs. The right illustrates the mismatching attack in that the VR user punches a virtual zombie overlay on a bystander. (b) Physical harm was identified in all collected scenarios, from mild pain or discomfort to extreme cases like death.}
\label{fig:vppm}
\Description{Figure 4a shows the results of the speculative design workshop. Participants were asked to create scenarios, where malicious actors exploit perceptual manipulations in VR to provoke physical harm to the user. I identified the puppetry attack, which directs the user's body action to provoke physical harm, and the mismatching attack, which makes the user misinterpret their virtual and physical information. Figure 4b shows the severity of the physical harm collected in the workshop. The harm varies from little pain and discomfort (e.g., eyestrain, cramp) to extreme harm (e.g., drowning, getting driven over).}
\end{center}
\end{figure*}

VR interaction does not have to stay in a one-to-one mapping (e.g., FingerMapper). Figure \ref{fig:thesis-structure} shows VR users perceive sensory information from the virtual and physical environments, and the perceptual input is dominated by the visual content in VR. Human-Computer Interaction (HCI) and VR research leverages this discrepancy to create VPPMs, effecting changes in the user's physical movements, becoming able to (perceptibly and imperceptibly) nudge their physical actions to enhance interactivity in VR. This project \cite{tsengDarkSidePerceptual2022} explores the risks of how malicious actors could provoke future physical breakdowns could be by exploiting VPPMs to manipulate a VR user's body motions (RQ1) and how to mitigate them (RQ2).

I chose speculative design workshop \cite{augerSpeculativeDesignCrafting2013, markussenPoeticsDesignFiction2013} to explore the future physical breakdowns and risks posed by the malicious use of VPPMs. The goal was to critique current practices and reflect on future technologies and their ethical implications. Through a thematic analysis \cite{braunUsingThematicAnalysis2006}, I analyzed 19 scenarios (Figure \ref{fig:vppm}a) created by eight VR and design experts, identifying two main risks and characterizing harm (Figure \ref{fig:vppm}b). The puppetry attacks control the physical actions of different body parts of an immersed user. The mismatching attacks are manipulations in which the adversary exploits a difference in information between a virtual object and its physical counterpart to elicit misinterpretation for the VR user. Two sample applications were implemented to show how existing VPPMs could be trivially subverted to create the potential for physical harm.Finally, I proposed mitigations and preventative recommendations against the malicious use of VPPMs. Unlike software leaks, one cannot patch a human perception hack easily. My goal was to raise awareness that the current way we apply and publish VPPMs can lead to malicious exploits of our perceptual vulnerabilities. 

\section{Future Work and Open Challenges}
In my future research, RQ3 is going to be the main focus. I plan to conduct experiments to understand the VR user's behavior and cognitive process of the physical breakdown phenomenon. One ongoing project is about the intentional break-outs and the concept of the Spatial Mental Model (SMM) --- a mental model that allows us to remember the spatial relation inside a small or well-known environment \cite{franklinSwitchingPointsView1992,  leeInterplayVisualSpatial2005}. Users can form an SMM of their physical environment (e.g., their home) and recall it despite only seeing the virtual environment while using VR. 

When an intentional break-outs happens, I assume the VR user might use their SMM of the physical and virtual environment to determine their interaction. Future research should focus on establishing this link even further to understand how these two constructs are related and what the user relies more on when breaking out. The goal would be identifying this process with SMM exists (i.e., the VR user not only using their perceptual information but their SMM in the physical environment to interact). Next, I want to observe the relationship between the user's SMM and other VR metrics (e.g., presence and enjoyment). An example could be a higher presence may lead to an imprecise SMM. My future research should create hypotheses around this direction and test them.

Physical breakdowns also represent a risk when using VR/XR. Future work could quantify physical breakdowns (or actions) and develop algorithms to predict the user's behavior. This approach can be a part of future safety mechanisms. I co-organized a workshop \cite{gugenheimerNovelChallengesSafety2022} focusing on the intersection of safety, security, and privacy in XR. Hacking human perception in VR (e.g., FruitSlicer or VPPMs) is also a potential risk in XR. I also plan to explore the countermeasure against manipulating human users with XR technology.

Most HCI and VR research inherit the concept of enhancing the VR experience (e.g., presence, immersion, and enjoyment) in a controlled environment. However, the context of everyday VR usage is uncertain and uncontrolled. Users may encounter multiple attentional transitions in one VR experience, switching between different realities and activities. One insight gained from my current progress is that applications should not only optimize for the user experience in VR. Instead, new metrics (e.g., safety) should be considered and enhanced in a different context. My future research aims to understand the cognitive process of how a user comprehends and interacts with the physical environment while being in VR. By investigating physical breakdowns in VR, I anticipate my thesis can contribute to the knowledge of VR users' behavior and cognitive process, further providing insights into future VR safety mechanisms.


\begin{acks}
I am grateful to my supervisors, Jan Gugenheimer, Samuel Huron, Eric Lecolinet, and my colleagues for their valuable feedback on my research. This work was conducted within INFRES and SES at Telecom Paris, IP Paris, LTCI, Institut Mines-Télécom, and the University of Glasgow. This work was supported by the Investments for the Future Program (PIA), CONTINUUM (ANR-21-ESRE-0030), and the INTERPLAY project (ANR-21-CE33-0022), funded by French National Research Agency (ANR).

\end{acks}

\bibliographystyle{ACM-Reference-Format}
\bibliography{doc-cons}

\end{document}